\documentclass[prl,a4paper,notitlepage,twocolumn,superscriptaddress,longbibliography,reprint]{revtex4-2}
\pdfoutput=1

\usepackage[utf8]{inputenc}
\usepackage{braket}
\usepackage{amsmath}
\usepackage{mathtools}
\usepackage{amssymb}
\usepackage{amsfonts}
\usepackage{graphicx}
\usepackage{hyperref}
\usepackage{comment}
\usepackage{color}
\usepackage{makecell}
\usepackage{enumitem}
\usepackage{upgreek}
\usepackage{hhline}
\usepackage{xfrac}

\usepackage{amsmath,amsthm,amssymb}
\usepackage[dvipsnames,table]{xcolor}
\usepackage{braket}
\usepackage{longtable}
\usepackage[bb=boondox]{mathalfa}
\usepackage{adjustbox}
\usepackage{array}
\usepackage{multirow}
\usepackage{tabularx}

\usepackage{float}

\usepackage{tikz}
\usetikzlibrary{shapes,arrows,patterns}

\newcommand{\X}{\mathrm{x}}

\newcommand{\Z}{\mathrm{z}}

\newcommand{\ccaption}[2]{\caption{\emph{#1.} #2}}
\newcommand{\ii}{\mathrm{i}}
\newcommand{\ie}{{\it i.e.},\ }
\newcommand{\eg}{{\it e.g.},\ }

\newcommand{\id}{\mathbb{1}} 

\newcommand{\Tr}{\operatorname{Tr}}
\newcommand{\m}{\operatorname{\gamma}}

\newcommand{\mysection}[1]{{\vspace{10 pt}\noindent \emph{{\textbf{#1}}.}}}

\usetikzlibrary{decorations.markings,arrows,decorations.pathreplacing,calc,intersections,through,backgrounds}

\tikzset{->-/.style={decoration={markings,mark=at position #1 with {\arrow{>}}},postaction={decorate}}}
\tikzset{
	partial ellipse/.style args={#1:#2:#3}{
		insert path={+ (#1:#3) arc (#1:#2:#3)}
	}
}
\begin{document}

\title{Long-distance entanglement of purification and reflected entropy \\in conformal field theory}

\author{Hugo A. Camargo}
\email{hugo.camargo@aei.mpg.de}
\affiliation{Max-Planck-Institut f\"{u}r Gravitationsphysik, Am M\"{u}hlenberg 1, 14476 Potsdam-Golm, Germany}
\affiliation{Dahlem Center for Complex Quantum Systems, Freie Universit\"{a}t Berlin, Arnimallee 14, 14195 Berlin, Germany}
\author{Lucas Hackl}
\email{lucas.hackl@unimelb.edu.au}
\affiliation{School of Mathematics and Statistics \& School of Physics, The University of Melbourne, Parkville, VIC 3010, Australia}
\author{Michal P. Heller}
\email{michal.p.heller@aei.mpg.de}
\altaffiliation[\emph{On leave of absence from:}]{ National Centre for Nuclear Research, Pasteura 7, 02-093 Warsaw, Poland}
\affiliation{Max-Planck-Institut f\"{u}r Gravitationsphysik, Am M\"{u}hlenberg 1, 14476 Potsdam-Golm, Germany}
\author{Alexander Jahn}
\email{a.jahn@fu-berlin.de}
\affiliation{Dahlem Center for Complex Quantum Systems, Freie Universit\"{a}t Berlin, Arnimallee 14, 14195 Berlin, Germany}
\affiliation{Institute for Quantum Information and Matter, California Institute of Technology, Pasadena, CA 91125, USA}
\author{Bennet Windt}
\email{bennet.windt17@imperial.ac.uk\\}
\affiliation{Blackett Laboratory, Imperial College London, Prince Consort Road, SW7 2AZ, UK}

\begin{abstract}
Quantifying entanglement properties of mixed states in quantum field theory via entanglement of purification and reflected entropy is a new and challenging subject. In this work, we study both quantities for two spherical subregions far away from each other in the vacuum of a conformal field theory in any number of dimensions. Using lattice techniques, we find an elementary proof that the decay of both, the entanglement of purification and reflected entropy, is enhanced with respect to the mutual information behaviour by a logarithm of the distance between the subregions. In the case of the Ising spin chain at criticality and the related free fermion conformal field theory, we compute also the overall coefficients numerically for the both quantities of interest.
\end{abstract}

\maketitle

\mysection{Introduction} Understanding quantum information properties of quantum field theory (QFT) and, through holography~\cite{Maldacena:1997re,Gubser:1998bc,Witten:1998qj}, also of gravity has been an important contemporary line of research~\cite{Casini:2009sr,Harlow:2014yka,Rangamani:2016dms,Susskind:2018pmk,Headrick:2019eth}. The main object of interest has been the entanglement entropy (EE) which reliably quantifies pure state entanglement between a subregion $A$ and its complement $\bar{A}$.
Given a reduced density matrix $\rho_{A} = \mathrm{tr}_{\bar{A}}\, \rho$ for a total pure state with density matrix $\rho$, EE is defined as the von Neumann entropy
\begin{equation}\label{eq:VNEntropy}
S_A=S(\rho_A) \equiv -\mathrm{tr}_{A} \rho_{A} \log{\rho_{A}}\,.
\end{equation}
EE is an ultraviolet-divergent quantity due to correlations at arbitrarily short distances in QFT and requires a regulator.
Efficient computations are possible using Gaussian techniques~\cite{sorkin1983entropy,peschel2003calculation,Weedbrook2012,Bianchi:2015fra,Hackl:2020ken} for free QFTs, analytical continuation methods for two-dimensional conformal field theory (CFT)~\cite{Holzhey:1994we,Calabrese:2004eu,Calabrese:2009qy,Calabrese:2009ez,Cardy:2013nua,Ugajin:2016opf}, or tensor network constructions for both gapped and gapless two-dimensional systems~\cite{Hastings:2007iok,Vidal:2008zz}. 
In strongly-coupled holographic QFTs, computing EE reduces to a geometric problem of finding minimal surfaces~\cite{Ryu:2006bv,Hubeny:2007xt,Lewkowycz:2013nqa,Dong:2016hjy}.

\begin{figure}
    \centering
    \includegraphics[width=0.45\textwidth]{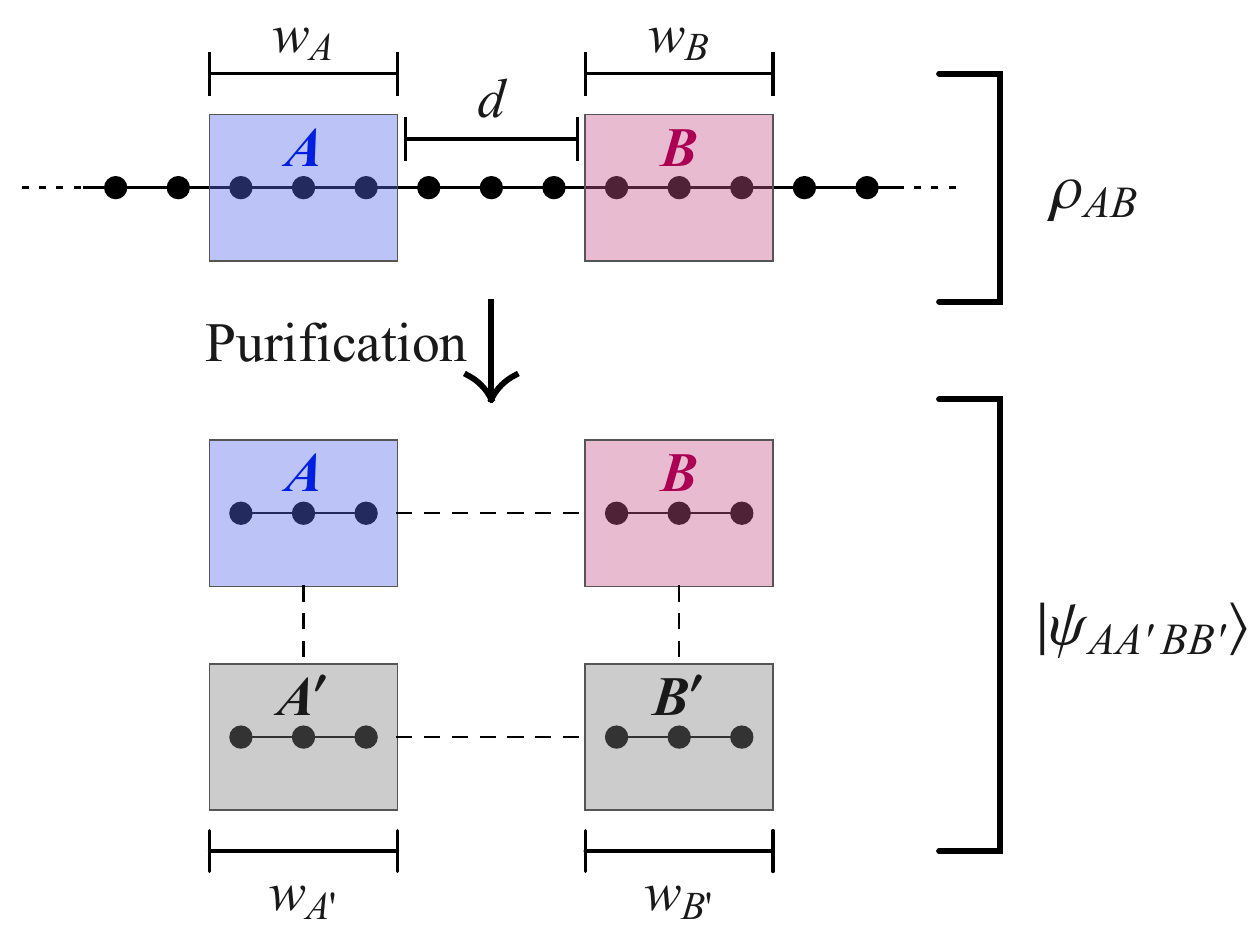}
    \caption{Illustration of our general setup for CFTs in two spacetime dimensions on a lattice: The mixed state $\rho_{AB}$ on a subsystem of two disjoint regions $A B$ separated by $N_{d} \equiv \frac{d}{\delta}$ sites is purified to a state with auxiliary factors $A^\prime$ and $B^\prime$, taken to be of the same size $N_{A} \equiv w_A/\delta$ and $N_{B} \equiv w_B/\delta$ as $A$ and $B$, respectively, where $\delta$ is the lattice spacing. Here we mostly consider $N_A=N_B$.
    }
    \label{fig:eop_def}
\end{figure}

In the present work, we will be concerned with CFTs in arbitrary number of dimensions emerging as a long-distance limit of lattice models regulated by a lattice spacing $\delta$. We will be interested in entanglement for subsystems composed of two disjoint regions $A$ and $B$, for which one often considers the \emph{mutual information} (MI) defined as
\begin{equation}
\label{eq.IAB}
I(A:B) = S_A + S_B - S_{A B}.
\end{equation}
A quantity of significant recent interest in such a setup is also the \emph{entanglement of purification} (EoP)~\cite{Terhal:2002}, which can be regarded as a generalization of EE for bipartite mixed states. It requires purifying the reduced density matrix $\rho_{AB}$ to a pure state $|\psi\rangle$ in an enlarged Hilbert space on $\mathcal{H}_{AB} \to \mathcal{H}_{A A^\prime B B^\prime}$ such that $\rho_{AB} = \mathrm{tr}_{A'B'} |\psi \rangle \langle \psi |$ (visualized in Fig.~\ref{fig:eop_def}). The EoP is then defined as
\begin{equation}
\label{eq.EoPdef}
E_{P}(\rho_{A B}) = \min_{\psi}[S_{A A'}].
\end{equation}
EoP is challenging to compute in QFT due to its inherent optimization procedure of finding a purification whose EE is minimal.
Its current understanding in the intersection of quantum information and high-energy physics is based on Gaussian calculations~\cite{Bhattacharyya:2018sbw,Bhattacharyya:2019tsi,Camargo:2020yfv}, CFT techniques with a limited range of applicability~\cite{Hirai:2018jwy,Caputa:2018xuf,Guo:2019pfl}, and on a conjectured realization in holography~\cite{Takayanagi:2017knl,Nguyen:2017yqw}. In the latter case, EoP has been conjectured to be dual to the entanglement wedge cross section~\cite{Czech:2012bh,Wall:2012uf,Headrick:2014cta,Bao:2017nhh}. This led to many novel developments regarding the emergence of the gravitational hologram~\cite{Dong:2016eik}, see~\eg~\cite{Umemoto:2018jpc,Tamaoka:2018ned,Bao:2018gck,Yang:2018gfq,Nomura:2018kji,Espindola:2018ozt,Liu:2019qje,Harper:2019lff,BabaeiVelni:2019pkw,Amrahi:2020jqg,Jain:2020rbb}. 

Another quantity closely related to EoP and also conjectured to be holographically dual to the entanglement wedge cross section is the \emph{reflected entropy}~RE~\cite{BabaeiVelni:2019pkw,Bao:2019zqc,Chu:2019etd,Dutta:2019gen,Kusuki:2019zsp,Jeong:2019xdr,Bueno:2020vnx,Bueno:2020fle,Zou:2020bly,Li:2020ceg,Kudler-Flam:2020url}. It is defined as EE
\begin{equation}
\label{eq.SRdef}
S_{R}(\rho_{AB}) = S_{AA'}(\ket{\sqrt{\rho_{AB}}}) 
\end{equation}
of the unique purification $\ket{\sqrt{\rho_{AB}}}:=\sum_i \sqrt{e_i}\ket{e_i}\ket{e_i}\in\mathcal{H}_A\otimes\mathcal{H}_B\otimes\mathcal{H}_{A'}\otimes\mathcal{H}_{B'}$ of $\rho_{AB}$, where  $\rho_{AB}\ket{e_i}=e_i\ket{e_i}$, $\mathcal{H}_{A'}=\mathcal{H}_{A}$ and $\mathcal{H}_{B}=\mathcal{H}_{B'}$. Simply put, $\ket{\sqrt{\rho_{AB}}}$ is the unique purification symmetric under $A\leftrightarrow A'$ and $B\leftrightarrow B'$ exchange. Clearly, $\ket{\sqrt{\rho_{AB}}}$ is one of the valid purifications $\ket{\psi}$ we minimize over in~\eqref{eq.EoPdef}, which implies $E_P\leq S_R$. RE is much easier to compute compared to EoP, as it does not require an optimization over all possible purifications.

The aim of this letter is to to elucidate a particularly simple setting in which EoP and RE behave universally across CFTs, without relying on Gaussianity or Weyl rescalings. We achieve this by using spin chains and more general lattice models and focusing on universal inequalities satisfied by EE. We corroborate our studies using analytics and numerics in the Ising and free fermion CFTs~\cite{DiFrancesco:1997nk}, which allows us to extract prefactors in the asymptotic scaling of EoP and~RE.

\mysection{Setup} In our analysis, we will be concerned with CFTs on a lattice. Our general statements will be made in any number of dimensions, whereas our numerics will focus on CFTs in two spacetime dimensions.

The setting of interest will contain two spherical subregions of diameter $w$ separated by a distance $d$. Fig.~\ref{fig:eop_def} illustrates it for CFTs in two spacetime dimensions in which case the subregions become intervals. At large distances $\frac{d}{w} \gg 1$, the decay of MI~\eqref{eq.IAB} in CFTs reads
\begin{equation}
\label{eq.MIcftlarged}
I(A:B) = {\cal N}\frac{\Gamma(\frac{3}{2})\Gamma(2\Delta + 1)}{2^{4\Delta +1} \Gamma(2\Delta + \frac{3}{2})} \times \epsilon_{\Delta}^2 + \ldots,
\end{equation}
where
\begin{equation}
\epsilon_{\Delta} \equiv \left( \frac{w}{d} \right)^{2 \Delta}
\end{equation}
and $\Delta$ corresponds of the scaling dimension of the lowest non-trivial operator(s) in the theory, $\cal N$ denotes the possible degeneracy of such operators and the ellipsis denotes faster decaying terms~\cite{Calabrese:2010he,Cardy:2013nua,Agon:2015ftl,Ugajin:2016opf}. The formula~\eqref{eq.MIcftlarged} assumes a gap in the spectrum of scaling dimensions and the lowest lying operator(s) being scalar(s). We will carry over this assumption in our studies of EoP and RE.

Our aim is to find and prove an analogue of the scaling in~\eqref{eq.MIcftlarged} for EoP and RE. In the latter case, recent numerical studies in free CFTs in~\cite{Bueno:2020vnx,Bueno:2020fle} led to the following fit
\begin{equation}
\label{eq.SRfit}
S_{R} =\alpha \, \epsilon_\Delta^{2} \log(\epsilon_\Delta^{-2})+\dots \quad\text{for}\quad \epsilon_{\Delta} \ll 1 \,.
\end{equation}
where $\alpha$ is a positive model-dependent constant.

In this letter, we use quantum-many body techniques in conjunction with elementary EE inequalities to \emph{prove} that the asymptotic form~\eqref{eq.SRfit} holds both for EoP and RE in a general CFT amenable to a lattice realization. 


\mysection{Elementary proof of the large distance behaviour} To set up the general argument valid both for EoP and RE, we only need to assume that the density operator $\rho_{AB}$ of two subsystems $A$ and $B$ far away from each other takes the form
\begin{align}
    \rho_{AB}(\epsilon_{\Delta})=\rho_{A}^{(0)}\otimes\rho^{(0)}_{B}+\epsilon_{\Delta} \, \rho_{AB}^{(1)}+\tfrac{1}{2}\epsilon_{\Delta}^2\rho^{(2)}_{AB} + \ldots\,,\label{eq.rhogeneral}
\end{align}
where the ellipsis denotes higher, not necessarily integer powers of $\epsilon_{\Delta}$ and we do not make any assumptions about subsystem sizes. The $\epsilon_\Delta$ term in~\eqref{eq.rhogeneral} is needed to reproduce the power-law scaling of correlation functions involving insertions of the lowest lying scaling operator in both $A$ and $B$. As we will show, the $\rho_{AB}^{(2)}$~contribution turns out to not contribute to the leading order decay of EoP and RE.

In the following, we will regard~\eqref{eq.rhogeneral} as originating from a perturbative purification
\begin{align}
\label{eq.defpertpuri}
    \ket{\psi}=\ket{\psi^{(0)}}+\epsilon_\Delta \ket{\psi^{(1)}}+\tfrac{1}{2}\epsilon^2_\Delta \ket{\psi^{(2)}}+\dots\,,
\end{align}
where the product nature of the density matrix~\eqref{eq.rhogeneral} for an infinite separation leads to
\begin{equation}
\label{eq.defpsi0}
\ket{\psi^{(0)}}=\ket{\psi^{(0)}_{AA'}}\otimes\ket{\psi^{(0)}_{BB'}}.
\end{equation}
Note that in our conventions, $|\psi\rangle$ and, therefore, also $|\psi^{(0)}\rangle$ are normalized, which also leads to constraint for $|\psi^{(j\geq 1)}\rangle$. The long distance behaviour of EoP and~RE is determined by the small-$\epsilon_{\Delta}$ expansion of the eigenvalues~$\mu_{j}$~of
\begin{equation}
\rho_{AA'} \equiv \mathrm{tr}_{BB'} |\psi\rangle \langle \psi|
\end{equation}
via the definition of EE~\eqref{eq:VNEntropy}: $S_{AA'}(\ket{\psi})=-\sum_{j\geq 0}\mu_j\log{\mu_j}$.

The fact that $\rho_{AB}$ is a product state for $\epsilon_{\Delta}=0$ implies that $\rho_{AA'}(\epsilon_{\Delta}=0)$ 
is itself pure, see~\eqref{eq.defpsi0}, and thus has eigenvalues $\mu_{0} = 1$ and $\mu_{j>0} = 0$. This result gets modified at large but finite distances. 

The linear correction to $\mu_j$ vanishes, since we expect $\mu_j$ to originate from a well-defined density matrix regardless of the sign of $\epsilon_{\Delta}$ when viewed as a formal parameter. As a result, the \emph{possible} leading behaviour of eigenvalues of~$\rho_{AA'}$ is given by
\small
\begin{align}
\mu_{0}\sim 1-\alpha_{\mathrm{tot}} \epsilon_{\Delta}^2\quad\text{and}\quad \mu_{j>0}\sim \alpha_j\epsilon_{\Delta}^2\quad\text{as}\quad\epsilon_\Delta\to0,\label{eq:mu-asymp}
\end{align}
\normalsize
where 
\begin{equation}
\label{eq.defalphatot}
\alpha_{\text{tot}} \equiv \sum_{j>0} \alpha_{j}.
\end{equation}
Note that $\alpha_{j>0} \geq 0$ and if all of them vanished, the behaviour encapsulated by~\eqref{eq:mu-asymp} would simply involve a higher-than-two power of $\epsilon_{\Delta}$.

Let us consider now the asymptotics of EoP and RE resulting from~\eqref{eq:mu-asymp}. As we explained in the introduction, these quantities are given by $S_{AA'}$ subject to additional conditions on purifications. For \emph{any} purification leading to~\eqref{eq:mu-asymp}, $S_{AA'}$ behaves as
\begin{equation}
\label{eq.SAAp}
S_{AA'} = \alpha_{\mathrm{tot}} \, \epsilon_{\Delta}^2 \, \log{\epsilon_{\Delta}^{-2}}+\beta \, \epsilon_{\Delta}^{2} + \ldots,
\end{equation}
where
\begin{equation}
\beta \equiv \Big( \sum_{j>0}\alpha_j(1-\log{\alpha_j}) \Big)
\end{equation}
and one sees as the leading order behaviour the structure~\eqref{eq.SRfit} identified in fits to free CFTs RE numerics in~\cite{Bueno:2020vnx,Bueno:2020fle} and the ellipsis denotes higher order terms~in~$\epsilon_{\Delta}$. 

Regardless of purification and long-distance limit, $S_{AA'}$ is bounded from below~\footnote{This serves as a lower bound for EoP and RE. However, for RE we even have the stronger bound $S_R\geq I_{AB}$ as shown in~(2.17) of~\cite{Dutta:2019gen}.}
\begin{align}
    S_{AA'}(\ket{\psi})\geq \frac{1}{2}I_{AB}(\rho_{AB})\,,\label{eq:ineq-bound}
\end{align}
as was shown in equation~(6) of~\cite{bagchi2015monogamy}. Given~\eqref{eq.MIcftlarged}, in order for~\eqref{eq:ineq-bound} to be satisfied at large distances $S_{AA'}$ cannot scale with a higher power than $\epsilon_{\Delta}^2$. Since the eigenvalue analysis predicts this as the strongest possible power-law factor in the long-distance behaviour of $S_{AA'}$, $\alpha_{\mathrm{tot}}$ must be bigger than~$0$ and the behaviour predicted by~\eqref{eq.SAAp} \emph{is necessarily} the behaviour of both EoP and RE in any CFT with a gap in the operator spectrum and amenable to a lattice description.

As a corollary of this proof, from the definition of~$\alpha_{\mathrm{tot}}$ in~\eqref{eq.defalphatot} we necessarily obtain that at least one of $\alpha_{j>0} > 0$ and, as a result, the first subleading term encapsulated in~\eqref{eq.SAAp} is also generically there. This is consistent with the findings of~\cite{Bueno:2020vnx,Bueno:2020fle}, which also identified such a contribution in RE for free CFTs on a lattice.

Finally, let us emphasize that our proof of the long-distance behaviour of EoP and RE did not rely on dimensionality of a CFT in question.

\mysection{Properties of the overall coefficient} Our proof predicts only that the overall prefactor $\alpha_{\mathrm{tot}}$ is positive. It is possible, however, to extract more information about what ingredients affect the exact value of $\alpha_{\mathrm{tot}}$ using a~rather general argument. To this end, notice that perhaps the easiest way to compute $\alpha_{\text{tot}}$ is to extract it from
\begin{align}
    \mathrm{Tr}(\rho_{AA'}^2)=1-2\alpha_{\mathrm{tot}}\epsilon_{\Delta}^2 + \ldots,\label{eq:TrrhoAA}
\end{align}
where we suppressed higher order terms in~$\epsilon_{\Delta}$.

Starting with~\eqref{eq.defpertpuri} and defining \begin{equation}
\label{eq.psiAApdef}
    \ket{\psi^{(i)}_{AA'}}=(\id\otimes\bra{\psi^{(0)}_{BB'}})\ket{\psi^{(i)}},
\end{equation}
we can write the reduction $\rho_{AA'}$ as
\begin{widetext}
\small
 \begin{align}
     \rho_{AA'}=\ket{\psi^{(0)}_{AA'}}\bra{\psi^{(0)}_{AA'}}+\epsilon_\Delta(\ket{\psi^{(0)}_{AA'}}\bra{\psi^{(1)}_{AA'}}+\ket{\psi^{(1)}_{AA'}}\bra{\psi^{(0)}_{AA'}})+\tfrac{1}{2}\epsilon_\Delta^2(2\mathrm{Tr}_{BB'}\ket{\psi^{(1)}}\bra{\psi^{(1)}}+\ket{\psi^{(2)}_{AA'}}\bra{\psi^{(0)}_{AA'}}+\ket{\psi^{(0)}_{AA'}}\bra{\psi^{(2)}_{AA'}}),
\end{align}
\normalsize
\end{widetext}
which allows us to compute $\mathrm{Tr}(\rho_{AA'}^2)$ explicitly. Upon using the normalization condition $\langle \psi | \psi \rangle = 1$ in the form of the following constraints
\begin{subequations}
\label{eq:constraints}
\begin{eqnarray}
    \braket{\psi^{(0)}|\psi^{(1)}}+\braket{\psi^{(1)}|\psi^{(0)}}&=&0\,, \label{eq:constraint1}\\
    \braket{\psi^{(0)}|\psi^{(2)}}+\braket{\psi^{(2)}|\psi^{(0)}}+2\braket{\psi^{(1)}|\psi^{(1)}}&=&0\,,\label{eq:constraint2}
\end{eqnarray}
\end{subequations}
we obtain
\small
\begin{align}
    \hspace{-10pt}\alpha_{\mathrm{tot}}=\lVert\ket{\psi^{(1)}}\rVert^2+|\langle\psi^{(0)}|\psi^{(1)}\rangle|^2-\lVert\ket{\psi_{AA'}^{(1)}}\rVert^2-\lVert\ket{\psi_{BB'}^{(1)}}\rVert^2\,,\label{eq:alpha}
\end{align}
\normalsize
where, in analogy with~\eqref{eq.psiAApdef}, $\ket{\psi_{BB'}^{(1)}}\equiv(\bra{\psi^{(0)}_{AA'}}\otimes \id)\ket{\psi^{(1)}}$. 

Quite remarkably and as advertised below~\eqref{eq.rhogeneral}, the correction quadratic in $\epsilon_{\Delta}$ to the density matrix $\rho_{AB}$ does \emph{not} contribute to the leading order coefficient in the scaling of both EoP and~RE. This looks like a potentially useful insight for any attempt to fix $\alpha_{\text{tot}}$ for EoP and RE in terms of CFT data in an analogous manner to~\eqref{eq.MIcftlarged} for MI. Furthermore, obtaining the leading behaviour of the EoP amounts simply to minimizing a quadratic polynomial obtained from components of $|\psi^{(1)} \rangle$ subject to the constraint~\eqref{eq:constraint1} and the condition (with $\rho_{AB}^{(1)}$ from~\eqref{eq.rhogeneral})
\begin{align}
    \rho_{AB}^{(1)}=\mathrm{Tr}_{B'A'}(\ket{\psi^{(1)}}\bra{\psi^{(0)}}+\ket{\psi^{(0)}}\bra{\psi^{(1)}})\,,\label{eq:affine-lin}
\end{align}
which generally leads to affine-linear constraints on~$\ket{\psi^{(1)}}$. The fact that the minimum exists follows from the argument presented in the previous section. While $\rho_{AB}^{(2)}$ does not affect the leading large distance behaviour encapsulated by $\alpha_{\text{tot}}$, individual $\alpha_{j>0}$ do depend on it and, via~\eqref{eq.SAAp}, so does the coefficient in front of the subleading term quadratic in $\epsilon_{\Delta}$.

\begin{table*}
\centering
\renewcommand{\arraystretch}{1.45}
\begin{tabular}{@{} c @{\hspace{0.4cm}} r @{$\quad$} r @{$\quad$} r p{.3cm} r @{$\quad$}  r @{$\quad$} r @{}}
\toprule
& \multicolumn{3}{c}{\bf{Free Fermions (Gaussian)}} & & \multicolumn{3}{c}{\bf{Ising Spins (non-Gaussian)}} \\
 \cline{2-4} \cline{6-8}
 & coefficient $\alpha_{\mathrm{tot}}$   & offset $\beta$ & equation & & coefficient $\alpha_{\mathrm{tot}}$   & offset $\beta$ & equation\\
 \cline{2-4} \cline{6-8}
MI & $0$  & $\tfrac{\log\frac{\pi+2}{\pi-2}}{4\pi}\approx 0.120$ & \eqref{eq:MIFerm11} & & $0$ & $C^2 \left( \frac{4\pi^2}{\pi^2-4} + \frac{\pi}{2}\log\frac{4+4\pi+\pi^2}{4-4\pi+\pi^2} \right)  \approx 0.298$& \eqref{eq.MIanalytic}
\\
EoP & $\frac{1}{8+2\pi^2}\approx 0.036$  &  $\frac{\log{2e(8+2\pi^2)}}{8+2\pi^2} \approx 0.181$ &  \eqref{eq:eopFerm11}
&  & $\frac{4C^2\pi^4}{\pi^4-16}\approx 0.124$&   $0.440$ &
\eqref{EQ_EOP_SPIN_W1}
\\
RE & $\frac{1}{2\pi^{2}}\approx 0.051$ & $\frac{1+\log(4\pi^{2})}{2\pi^{2}} \approx 0.237$ & \eqref{EQ_RE_FERMION_W1}
 &  & $\frac{4C^{2}(\pi^{2}-2)}{\pi^{2}-4}\approx0.139$&   $0.425$& \eqref{EQ_RE_SPIN_W1}
\\[1mm]
\colrule
\botrule
\end{tabular}
\caption{Summary of numerical and analytical results for the leading coefficient $\alpha_{\mathrm{tot}}$ and the offset $\beta$ obtained for MI, EoP and RE with asymptotics \eqref{eq.SAAp} both for Ising spins and for latticized fermions on a line. We refer to the respective equation in the Supplemental Material. Numbers without analytical expression are based on a numerical fit.
}
\label{tab:Alpha_Offset}
\end{table*}

\mysection{Analysis in the critical Ising chain and free fermion CFT} So far, we have been completely general in our studies and in the following we will specialize to two closely related lattice models describing CFTs in two spacetime dimensions. This will allow us to obtain numerical values of the leading and first subleading coefficients in the behaviour of EoP and RE captured by~\eqref{eq.SAAp} and, for RE, compare with earlier studies in~\cite{Bueno:2020vnx}.

The Ising model realization of the $c = \frac{1}{2}$ CFT on an infinite spatial line can be described by the critical lattice Hamiltonian
\begin{equation}
\label{eq.HIsing}
\hat{H}\sim-\sum_{i = -\infty}^{\infty} (2\,\hat{S}^{\X}_i \hat{S}^{\X}_{i+1}+\,\hat{S}^{\Z}_i) \ ,
\end{equation}
more general forms of which we discuss in the Supplemental Material. The $\hat{S}^{\X,\Z}_{i}$ are spin operators defined by the Pauli matrices $\hat{S}^{\X,\Z}_{i}=\frac{1}{2}\sigma^{\X,\Z}_{i}$. In the Ising CFT there is a non-degenerate (\ie ${\cal N} = 1$) lightest operator of scaling dimension $\Delta = \sfrac{1}{8}$, often denoted as the spin field $\sigma$ and corresponding to a $\hat{S}_{i}^{\X}$ lattice operator.

The critical Ising model can be mapped to a free fermion theory, which is going to be another model in which we obtain numerical coefficients in EoP and RE. This formulation leads to two different notions of reduced density matrices for disjoint intervals, see, for example,~\cite{Igloi_2010,Fagotti:2010yr,Coser:2015mta,coser2016spin,Camargo:2020yfv}, and therefore provides in itself an independent example. For free fermion CFT there are two (${\cal N} = 2$) lowest lying operators with $\Delta = \sfrac{1}{2}$ and being simply the fermionic field operators.

\begin{figure*}[htb]
\centering
\begin{tabular}{c p{0.1cm}  c p{0.5cm} c}
    & & \hspace{0.5cm} {\bf Free fermions} & & \hspace{0.5cm}  {\bf Ising spins} 
    \\[0.15cm]
    \rotatebox{90}{\hspace{0.048\textheight} {\bf MI results}} & &
    \includegraphics[width=0.44\textwidth]{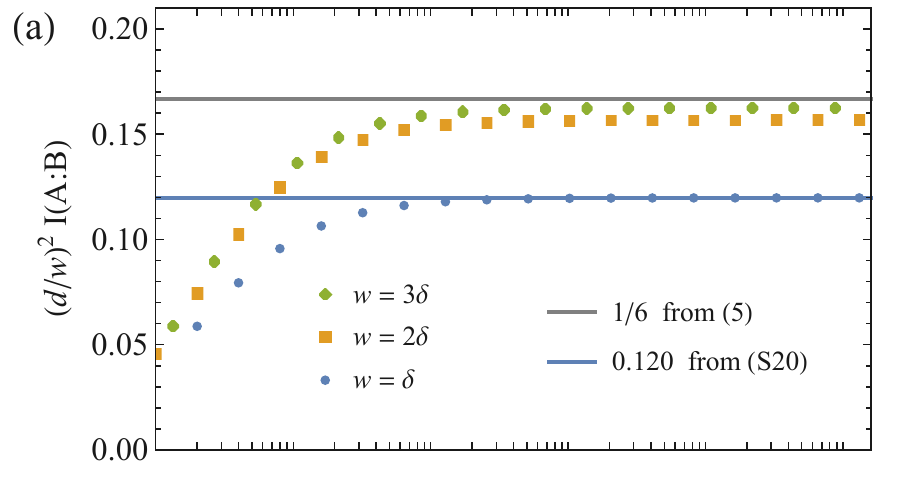} & &
    \includegraphics[width=0.44\textwidth]{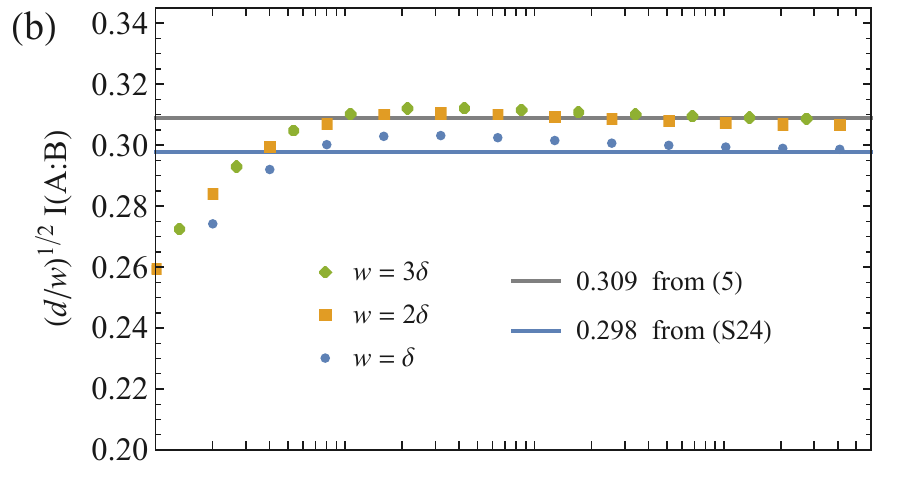} 
    \\[0.15cm] 
   \rotatebox{90}{\hspace{0.076\textheight} {\bf EoP results}} & & \includegraphics[width=0.44\textwidth]{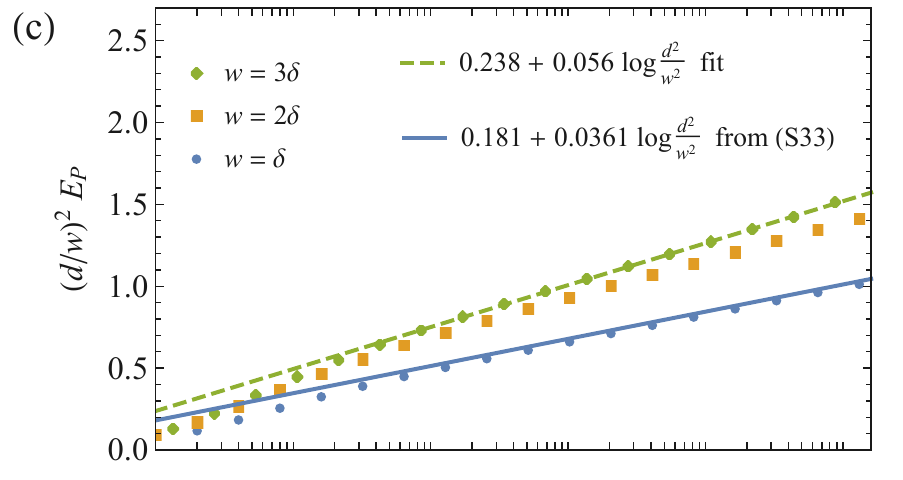} & &
    \includegraphics[width=0.44\textwidth]{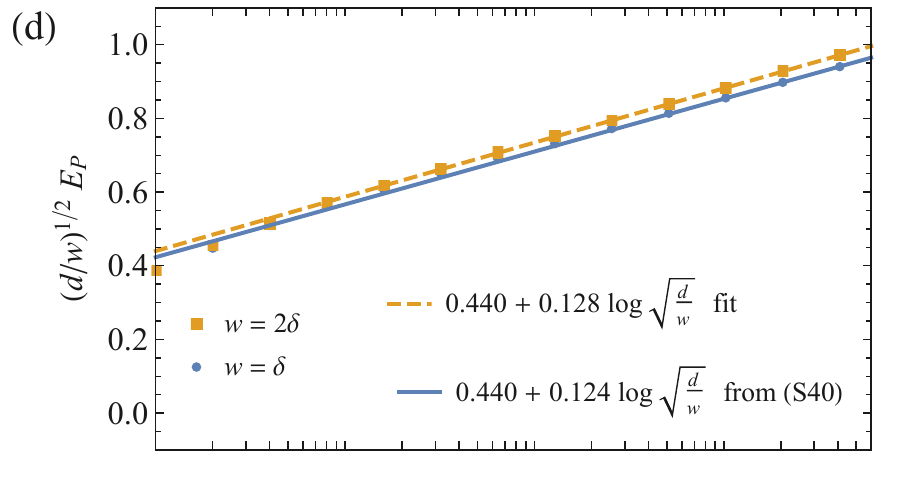} \\[0.15cm] 
   \rotatebox{90}{\hspace{0.082\textheight} {\bf RE results}} & & \includegraphics[width=0.44\textwidth]{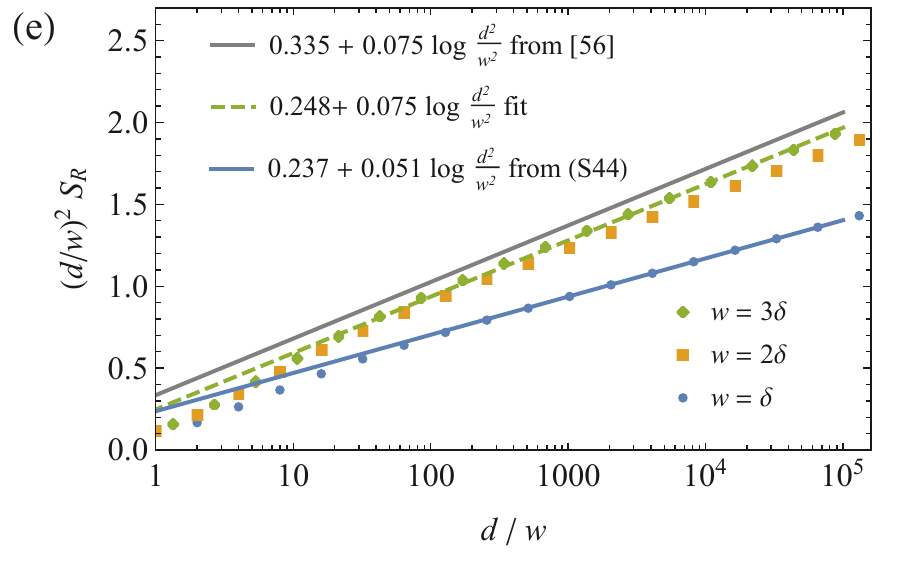} &
   &
    \includegraphics[width=0.44\textwidth]{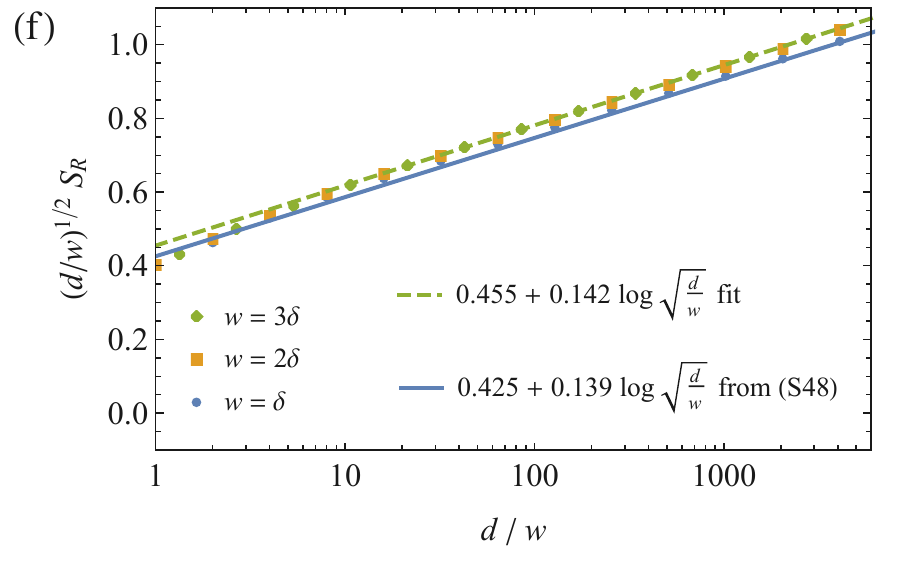}
\end{tabular}
\caption{Numerical data for MI, EoP and RE for fermions and spins, rescaled by the power-law contribution $\epsilon_\Delta^2 = (w/d)^{4\Delta}$ of the respective dominant term with $\Delta=\sfrac{1}{2}$ for free fermions and $\Delta=\sfrac{1}{8}$ for Ising spins. The analytical predictions for $w=\delta$ are derived in the Supplemental Material (see also Tab.~\ref{tab:Alpha_Offset}). Analytical comparisons are drawn as solid lines and fits of the numerical EoP and RE data at the largest available $w$ as dashed ones. The top solid (grey) line in (e) displayed above the numerical data corresponds to the result reported in~\cite{Bueno:2020vnx}.
}
\label{fig:MI_EoP}
\end{figure*}

Critical lattice models will describe CFT predictions for large enough sizes of subsystems at fixed~$w/d$. Since we are dealing with purifications, which can lead to challenges when the relevant Hilbert space dimension becomes big, the key question is how big subsystems need to be to reproduce the continuum physics of interest. One hint comes from MI, for which we see that the continuum value of the prefactor in~\eqref{eq.MIcftlarged} is well attained at large distances already for $w = 2\, \delta$ and $3 \, \delta$ with the smallest subsystems of $w = \delta$ giving already reasonable predictions, see Fig.~\ref{fig:MI_EoP}(a,b). 

Given this encouraging result, we were in fact able to \emph{analytically} compute the coefficients of the leading order MI, EoP and RE in the critical Ising model and for free fermions when $w = \delta$. The results are summarized in Tab.~\ref{tab:Alpha_Offset} together with offsets and derivations can be found in~Supplemental Material.

Fig.~\ref{fig:MI_EoP}(c-f) shows fits of our proven asymptotic formula~\eqref{eq.SAAp} to fully numerical results, \ie based on the full density matrix for disjoint intervals of both CFTs, we consider. We see strong indications of convergence to continuum values. In particular, for the critical Ising model EoP, the behaviour of $\alpha_{\text{tot}}$ for $w = \delta$ corroborates our analytical prediction in Tab.~\ref{tab:Alpha_Offset}. Looking at the results for the largest atteinable values of $w$, we see that the leading fall-off coefficient changes from the analytic prediction at $w = \delta$ by only $2.6\%$ and the subleading fall-off coefficient by only $3.9\%$. Generating data for $w = 3 \, \delta$ and above is numerically challenging as this requires computing very large matrices, slowing down calculations (see Supplemental Material).

For the EoP and RE of the Ising CFT, our results provide to the best of our knowledge new predictions, whereas for RE for the massless free fermions we find and display very good agreement with earlier studies in~\cite{Bueno:2020vnx}. 

An interesting question regarding EoP concerns the dimensions of the enlarged Hilbert spaces. In our setup, when purifying the state of a system with $N_A+N_B$ degrees of freedom by adding $N_{A'}+N_{B'}$ additional ones (see also Fig.~\ref{fig:eop_def}), there is \emph{a priori} no constraint on $N_{A'},N_{B'}$ other than the basic requirement following from the definition of the Schmidt decomposition that $N_{A'}+N_{B'}\geq N_A+N_B$. However, we show in the Supplemental Material that the choice of \textsl{minimal purifications} used so far yields the true minimum of EE as long as we choose $N_{A'}=N_A$ and $N_{B'}=N_B$. This was already shown in~\cite{Windt:2020tra} based on ideas of~\cite{Hackl:2020viw} for the EoP for Gaussian states (with Gaussian purifications).

\mysection{Outlook} In this letter, we studied large distance behaviour of EoP and RE in a generic CFT with a gap in the operator spectrum for two spherical subsystems of diameter $w$ in the large distance $d$ limit. Using~\eqref{eq.rhogeneral} in conjunction with elementary properties of EE we were able to \emph{prove} that the large order behaviour of both EoP and RE is governed by~\eqref{eq.SAAp}. In comparison to the classic result~\eqref{eq.MIcftlarged} encapsulating large-distance behaviour of MI, EoP and RE get both enhanced by a logarithm of a separation. Subsequently, we explicitly calculated the large distance behaviour for both EoP and RE in one spatial dimension for the critical Ising model and massless fermions. This allowed us to establish the value of coefficients appearing in~\eqref{eq.SAAp}, see Tab.~\ref{tab:Alpha_Offset} and Fig.~\ref{fig:MI_EoP}.

Our work opens a genuinely new avenue for studying EoP and RE in QFTs without restriction to free models. Perhaps the most interesting question concerns the dependence of the coefficients in the large order behaviour of EoP and RE on CFT data, akin to~\eqref{eq.MIcftlarged} for MI. An intermediate step could be to supplement our numerical code with large-distance reduced density matrices obtained with tensor networks for more complicated models, in particular determining model-dependent coefficients akin to~\eqref{eq.MIcftlarged}. Optimizing over purifications outside the Gaussian realm inevitably leads to vast parameter spaces that quickly exhaust desktop-scale computational resources. However, the entanglement between reasonably-sized subsystems both mixed and purified is not large and it should be possible to represent purifications as manageable tensor networks, perhaps building on earlier works~\cite{Nguyen:2017yqw,TNpurifications}.

\begin{acknowledgments}
\mysection{Acknowledgments}
We would like to thank J.~Eisert and T.~Takayanagi for collaborations on related subjects and M.~C.~Ba{\~n}uls, T.~Faulkner, J.~Knaute, C.~Pattison, D.~Radicevic, L.~Shaposhnik, S.~Singh, V.~Svensson, B.~Swingle and L.~Tagliacozzo for useful discussions and comments on the draft. Our special thanks go to P.~Bueno who in response to the first version of the manuscript pointed out to us that the behavior encapsulated by~\eqref{eq.SRfit} was also seen in the reflected entropy in free fermion and free boson QFTs~\cite{Bueno:2020vnx,Bueno:2020fle}. The Gravity, Quantum Fields and Information group at the Max Planck Institute for Gravitational Physics (Albert Einstein Institute) is supported by the Alexander von Humboldt Foundation and the Federal Ministry for Education and Research through the Sofja Kovalevskaja Award. AJ is supported by the FQXi. HC is partially supported by the Konrad-Adenauer-Stiftung through their Sponsorship Program for Foreign Students and by the International Max Planck Research School for Mathematical and Physical Aspects of Gravitation, Cosmology and Quantum Field Theory.
\end{acknowledgments}

\appendix


\appendix

 \setcounter{figure}{0}
 \setcounter{table}{0}
 \setcounter{equation}{0}



\section*{Appendix}

\renewcommand{\thetable}{S\arabic{table}}  
\renewcommand{\thefigure}{S\arabic{figure}} 
\renewcommand{\theequation}{S\arabic{equation}}



\mysection{Review of critical Ising model} The Hamiltonian of the transverse Ising model is given by
\begin{equation}\label{eq:TIsingHam}
\hat{H} = - \sum_{k=1}^N \left( 2 J\, \hat{S}^x_k \hat{S}^x_{k+1} + h\, \hat{S}^z_k \right) \ ,
\end{equation}
with spin operators represented by Pauli matrices $\sigma_\alpha$ with $\alpha \in (x,y,z)$ by 
\begin{equation}
\hat{S}^\alpha_k \equiv \id^{\otimes (k-1)} \otimes \frac{\sigma_\alpha}{2} \otimes \id^{\otimes (N-k)} \ .
\end{equation}
We also use the identification $\hat{S}^\alpha_{N+1} \equiv \hat{S}^\alpha_1$.
This spin model can be converted to fermions by defining the $2N$ Majorana operators $\m_k$ via
\begin{align}
\m_{2k-1} &= {\sigma_z}^{\otimes (k-1)} \otimes \sigma_x \otimes \id^{\otimes (N-k)} \ , \\
\m_{2k} &= {\sigma_z}^{\otimes (k-1)} \otimes \sigma_y \otimes \id^{\otimes (N-k)} \ .
\end{align}
The Ising Hamiltonian then takes the form
\small
\begin{equation}
\hspace{- 10 pt}\hat{H}_\text{I} = \frac{\ii}{2} \left( \m_1 \m_{2N} P + J \sum_{k=1}^{N-1} \m_{2k} \m_{2k+1} + h \sum_{k=1}^{N} \m_{2k-1} \m_{2k} \right) .
\end{equation}
\normalsize
Here $P$ is the total parity operator ${\sigma_z}^{\otimes N}=\prod_k (-\ii \m_{2k-1}\m_{2k})$.
At the critical point $J=h$, the Hamiltonian thus simplifies to
\begin{equation}
\hat{H} = \frac{\ii}{2} \left( \m_1 \m_{2N} P + \sum_{k=1}^{2N-1} \m_k \m_{k+1} \right)\,,
\end{equation}
which leads for $N\to\infty$ to the lattice model of the $c=\tfrac{1}{2}$ CFT. The critical Ising Hamiltonian as displayed in the main text~\eqref{eq.HIsing} 
corresponds to $J = h = 1$.

\mysection{Covariance matrix} For the critical ground state vector $\ket{0}$ which has a positive total parity, all correlations are encoded in the Majorana covariance matrix 
\begin{equation}
\Omega_{j,k} = \frac{\ii}{2} \bra{0} [\m_j, \m_k] \ket{0} \ ,
\end{equation}
which in the infinite system size limit takes the form
\begin{equation}
\Omega_{j,k} =
\begin{cases}
0 &  k = j  \\
\frac{(-1)^{k-j} - 1}{\pi (k-j)} &  k \neq j
\end{cases} \ .\label{eq:covariance-inf}
\end{equation}
The entropy of a Gaussian mixed state $\rho$ with covariance matrix $\Omega$ is given by
\begin{align}
    S(\rho)=-\sum_{\pm, i}\frac{1\pm\lambda_i}{2}\log\frac{1\pm\lambda_i}{2}\,,\label{eq:Gaussain-SA}
\end{align}
where $\pm\ii \lambda_k$ are the eigenvalues of $\Omega$. We will consider mixed states $\rho_{AB}$ or $\rho_{AA'}$, whose mixed state covariance matrices $\Omega_{AB}$ or $\Omega_{AA'}$ result from restricting~\eqref{eq:covariance-inf} to the respective blocks.

\mysection{Fermionic subsystem}
\begin{figure}
    \centering
    \includegraphics[width=0.46\textwidth]{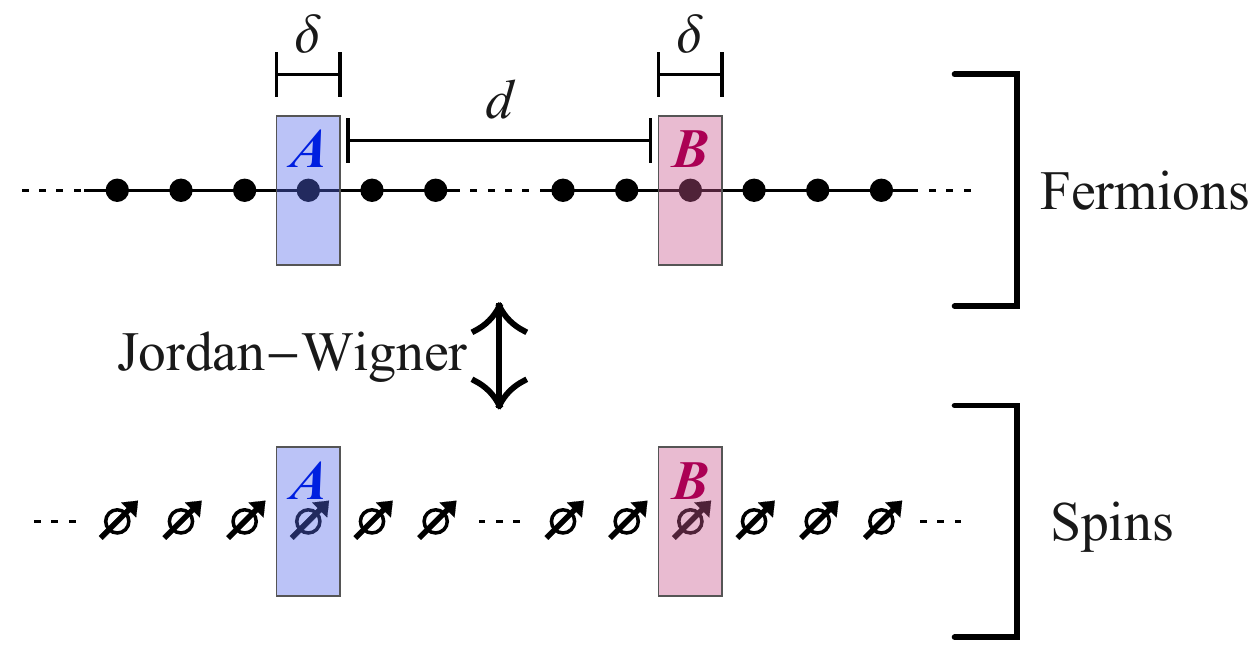}
    \caption{Subsystem setup of our analytical limits for fermions (top) with an inherent ordering and spins (bottom) without one. In both systems, we consider the subsystem $AB$ consisting of two single sites $A$ and $B$ separated by $d/\delta$ sites.
    }
    \label{fig:1+1subsystem}
\end{figure}
We first compute reduced density matrices from the perspective of fermions, \ie imposing an ordering between modes following from the anti-commuting variables $\gamma_i$ (see Fig.~\ref{fig:1+1subsystem}). A subsystem consisting of $1+1$ sites ($w=\delta$) separated by $d/w=d/\delta$ sites is then fully characterized by the restriction of the covariance matrix in~\eqref{eq:covariance-inf} and explicitly given by
\begin{align}
    \Omega^{\mathrm{fer}}_{AB} = 
    \begin{pmatrix}
      & -\frac{2}{\pi } &  & \frac{-2}{(2 d/w+3) \pi } \\
     \frac{2}{\pi } &  & \frac{-2}{(2 d/w+1) \pi } &  \\
      & \frac{2}{(2 d/w+1) \pi } &  & -\frac{2}{\pi } \\
     \frac{2}{(2 d/w+3) \pi } &  & \frac{2}{\pi } &  
    \end{pmatrix}\label{eq:omega-fermions}
\end{align}
which corresponds to a lowest-dimension (Majorana) operator with scaling dimension $\Delta=\sfrac{1}{2}$. The associated fermionic density operator is then
\begin{align}\label{eq:rho-fermions}
    \hspace{-2mm}\rho^{\mathrm{fer}}_{AB}\sim\begin{pmatrix}
     D & & &\frac{1}{2\pi}\,\epsilon_{\sfrac{1}{2}}\\
     & E & & \\
     & & E & \\
     \frac{1}{2\pi}\,\epsilon_{\sfrac{1}{2}} & & & F
    \end{pmatrix}
\end{align}
with respect to the basis $(\ket{\downarrow\downarrow}$, $\ket{\uparrow\downarrow},\ket{\downarrow\uparrow},\ket{\uparrow\uparrow})$ and using $D=\frac{1}{4}+\frac{1}{\pi}+\frac{1}{\pi^2}$, $E=\frac{1}{4}-\frac{1}{\pi^2}$, and $F=\frac{1}{4}-\frac{1}{\pi}+\frac{1}{\pi^2}$.
As in the main text, $\epsilon_\Delta \equiv (w/d)^{2\Delta}$ which here becomes $\epsilon_{\sfrac{1}{2}}=w/d$.
If we further restrict to a single site, we find the covariance matrix and density operator
\begin{align}
    \Omega^{\mathrm{fer}}_{A} &= 
    \begin{pmatrix}
      & -\frac{2}{\pi }  \\
     \frac{2}{\pi } &
    \end{pmatrix} & \rho_A^{\mathrm{fer}} &= 
    \begin{pmatrix}
     \frac{1}{2}-\frac{1}{\pi} & \\
      & \frac{1}{2}+\frac{1}{\pi}
    \end{pmatrix}\,,
\end{align}
where $\rho_A^{\mathrm{fer}}$ is written with respect to the basis $(\ket{\downarrow},\ket{\uparrow})$.

\mysection{Spin subsystem} We can perform a similar calculation in the original Ising spin system whose reduced density matrices can be constructed from the fermionic covariance matrix \cite{Coser:2015mta}. We need not repeat the single interval case, as entanglement entropies of connected regions are equivalent under a Jordan-Wigner transformation. However, we still need the reduced density matrix of a system of $1+1$ sites in the large $d$ limit, which we find to be
\begin{equation}
\label{eq:rhoAB}
    \rho^{\mathrm{spin}}_{AB} \sim 
    \begin{pmatrix}
      D &  &  & C\epsilon_{\sfrac{1}{8}} \\
  & E & C\epsilon_{\sfrac{1}{8}} & \\
  & C\epsilon_{\sfrac{1}{8}} & E &  \\
 C\epsilon_{\sfrac{1}{8}} &  &  & F
    \end{pmatrix}
\end{equation}
with $w/\delta=1$ for the setup of $1+1$ sites. As we are considering the spin Ising CFT, the lowest-dimension primary is the ``order field'' $\sigma$ with scaling dimension $\Delta=\sfrac{1}{8}$. 
The constant $C$ corresponds to the expectation value of an operator nonlocal in fermions, and can be computed from
\begin{equation}
    C = \lim_{n \to \infty} \left( \frac{2}{\pi} \right)^n \frac{n^{1/4}}{4} \det{M^n} \ ,
\end{equation}
where $M^n$ is defined as the $n \times n$ matrix
\begin{equation}
    M^n_{j,k} = 
    \begin{cases}
    \frac{(-1)^{k-j}}{2(k-j)+1} & j \leq k \\
    \frac{(-1)^{j-k+1}}{2(j-k)-1} & j > k
    \end{cases} \ .
\end{equation}
Using this construction, one finds~\cite{McCoy:2010}
\begin{equation}
\label{eq:Cdefinition}
    C =\frac{e^{3\zeta'(-1)}}{2^{23/12}}\approx 0.1612506\,.
\end{equation}

\mysection{MI for fermions}
We now begin computing entanglement measures for the small subsystems whose reduced density matrices we just obtained explicit expressions for.
The continuum limit corresponding to the Ising CFT is obtained by keeping $d/w$ (or, equivalently, $\epsilon_\Delta$) fixed and taking $\delta/w$ to $0$. We will see that taking only a few lattice sites is sufficient to describe the qualitative and approximate quantitative behavior of the continuum limit.
To demonstrate this, we now show that the large distance asymptotics of the MI ~\eqref{eq.IAB} for the case of $1+1$ sites yield results close to the continuum formula~\eqref{eq.MIcftlarged}. 
In order to compute the MI, we need to determine the von Neumann entropy of a single site $S_A=S_B$ and of both sites $S_{AB}$. These entropies can be computed from the eigenvalues of the covariance matrix $\Omega$ associated to the respective Gaussian state $\rho$. As an antisymmetric matrix, $\Omega^{\mathrm{fer}}_{AB}$  has pairs of purely imaginary eigenvalues $\pm \ii\, \lambda_k$, from which applying~\eqref{eq:Gaussain-SA} leads to
\begin{align}
S_A&=- \tfrac{\pi+2}{2\pi}\log\tfrac{\pi+2}{2\pi} - \tfrac{\pi-2}{2\pi}\log\tfrac{\pi-2}{2\pi} \approx 0.474\\
    S_{AB} &=\sum_{k=1}^n \left( - \tfrac{1+\lambda_k}{2}\log\tfrac{1+\lambda_k}{2} - \tfrac{1-\lambda_k}{2}\log\tfrac{1-\lambda_k}{2} \right)\,,\label{eq:Gaussian-SAB}
\end{align}
where the eigenvalues of $\Omega_{AB}^{\mathrm{fer}}$ are to leading order
\begin{equation}
    \lambda_{1,2} = \frac{1}{\pi} \left( 2 \pm \tfrac{3}{4}\epsilon_{\sfrac{1}{2}}^2  + \dots \right) \ . 
\end{equation}
We can similarly expand $S_{1+1} \equiv S_{AB}$ at large $d$, which results in a MI for $w=\delta$ of
\begin{align}\label{eq:MIFerm11}
    I^{\mathrm{fer}}(A:B)&\sim \tfrac{\log\frac{\pi+2}{\pi-2}}{4\pi}\,\epsilon_{\sfrac{1}{2}}^2 \nonumber\\
    &= 0.120 \,\left(\tfrac{w}{d}\right)^2 \ .
\end{align}
This reproduces the correct continuum power law of fermionic MI, but yields a coefficient lower than the continuum value \eqref{eq.MIcftlarged} which also matches the large-distance expansion of earlier results for Dirac fermions~\cite{Casini:2005rm}
\begin{align}\label{eq:MIFerm}
    I(A:B)=\frac{c}{3}\log\frac{(d+w)^2}{d\,(2w+d)} &\sim \frac{1}{6} \left(\frac{w}{d}\right)^2 \ ,
\end{align}
for two blocks of general width $w$.

\mysection{MI for spins} 
We compute EE for the spin system directly from the eigenvalue spectrum of the reduced density matrix \eqref{eq:rhoAB}. Its four eigenvalues $\mu_j$ are
\begin{subequations}
\begin{align}\label{eq:EigenSpinRhoAB}
    \mu_{1,2} &= \frac{1}{4} - \frac{1}{\pi} \pm C \,\epsilon_{\sfrac{1}{8}} \ , \\
    \mu_{3,4} &= \frac{1}{4} + \frac{1}{\pi} \pm \sqrt{\frac{1}{\pi^2} + C^2 \, \epsilon_{\sfrac{1}{8}}^2}\ ,
\end{align}
\end{subequations}
where $C$ is given by \eqref{eq:Cdefinition} and from which we can directly compute the EE for $AB$ via
\begin{equation}
\label{eq.Smu}
S = - \sum_j \mu_j \log\mu_j.
\end{equation}
Note that in the following we will denote eigenvalues of any density matrix by $\mu_{j}$.

This analysis leads to the Ising model prediction for spin MI at $w=\delta$ and large separations $d$ of the form
\begin{align}
\label{eq.MIanalytic}
 \hspace{-2mm}   I^\text{spin}(A:B) &\sim 
 C^2 \left( \frac{4\pi^2}{\pi^2-4} + \frac{\pi}{2}\log\frac{4+4\pi+\pi^2}{4-4\pi+\pi^2} \right)  \epsilon_{\sfrac{1}{8}}^2 \nonumber\\
 &\sim 0.298\, \sqrt{\frac{w}{d}} \ ,
\end{align}
\normalsize
Comparing this formula with the CFT analytics~\eqref{eq.MIcftlarged} for $\Delta = \sfrac{1}{8}$, we see an exact match in the power-law behavior. Furthermore, the prefactor in~\eqref{eq.MIanalytic} is only $3.6\%$ off from the continuum value $\approx0.309$ predicted by~\eqref{eq.MIcftlarged}.

\mysection{EoP for fermions} Analogous to the MI calculation for free fermions, we now calculate the EoP in the fermionic subsystem of two sites separated by $d/\delta$ sites, expressing all calculations in terms of covariance matrices. We purify $\Omega_{AB}$ in the limit $d/\delta\to\infty$ as
\begin{align}
    \Omega^{(0)}=\left(\begin{array}{cccc|cccc}
      & -G & & &  & L& & \\
    G &    & & & L &  & &\\
      &  & & -G & & & & L\\
     &    & G & & &  & L&\\
     \hline
       & -L & & &  & -G& & \\
    -L &    & & & G & & &\\
      &  & & -L & & &  & -G\\
     &    & -L & & &  & G&\\
    \end{array}\right)
\end{align}
associated to systems $(A,B,A',B')$ with $G=\frac{2}{\pi}$ and $L=\sqrt{1-G^2}$, whose EE $S_{AA'}$ is zero and we thus have $\lim_{d\to\infty}E_P=0$, \ie the EoP vanishes for large $d/\delta$, as expected.

In order to find the asymptotic behavior of $E_P$, we need to study the variation of the symplectic eigenvalues $\pm\ii\lambda_i$ of $\Omega_{AA'}$ when perturbing $\Omega$ according to
\begin{align}
    \Omega\sim \Omega^{(0)}+\epsilon_{\sfrac{1}{2}}\,\Omega^{(1)}+\tfrac{1}{2}\epsilon_{\sfrac{1}{2}}^2\,\Omega^{(2)}\quad\text{as}\quad \epsilon_{\sfrac{1}{2}}\to 0\,.
\end{align}
The requirement of $\Omega$ representing a purification implies $\Omega^2=-\id$, which induces the constraints
\begin{align}
\begin{split}
    \Omega^{(0)}\Omega^{(1)}+\Omega^{(1)}\Omega^{(0)}&=0\,,\\
    2(\Omega^{(1)})^2+\Omega^{(0)}\Omega^{(2)}+\Omega^{(2)}\Omega^{(0)}&=0\,.
\end{split}\label{eq:constraints-J}
\end{align}
We further require that the restrictions $\Omega^{(1)}_{AB}$ and $\Omega^{(2)}_{AB}$ matches the ones of\ \eqref{eq:omega-fermions} expanded in $\epsilon_{\sfrac{1}{2}}$, \ie
\begin{align}
    \Omega^{(1)}_{AB}\!=\!\begin{pmatrix}
    &&& -\frac{1}{\pi}\\
    &&-\frac{1}{\pi}& \\
    &\frac{1}{\pi}&& \\
    \frac{1}{\pi}&&&
    \end{pmatrix},\,\,
    \Omega^{(2)}_{AB}\!=\!\begin{pmatrix}
    &&& \frac{3}{\pi}\\
    &&\frac{1}{\pi}& \\
    &-\frac{1}{\pi}&& \\
    -\frac{3}{\pi}&&&
    \end{pmatrix}\,,\label{eq:constraints2-J}
\end{align}
The equations\ \eqref{eq:constraints-J} and\ \eqref{eq:constraints2-J} can be solved iteratively up to some free variables. We first solve $\Omega^{(1)}$ in terms of $\Omega^{(0)}$ and then $\Omega^{(2)}$ in terms of $\Omega^{(0)}$ and $\Omega^{(1)}$.

In order to find asymptotics of the symplectic eigenvalues $\lambda_i$, we can use the fact that $\Tr(\Omega_{AA'}^2)=-2(\lambda_1^2+\lambda_2^2)$ and $\Tr(\Omega_{AA'}^4)=2(\lambda_1^4+\lambda_2^4)$ to solve for the asymptotics of $\lambda_i$ to be given by
\begin{align}
    \lambda_1=\lambda_2\sim 1-\alpha_{\mathrm{tot}}\,\epsilon_{\sfrac{1}{2}}^2\quad \text{as}\quad \epsilon_{\sfrac{1}{2}}\to 0\,,\label{eq:symp-eig}
\end{align}
where $\alpha_{\mathrm{tot}}$ will depend on some of the free parameters contained in $\Omega^{(1)}$ and $\Omega^{(2)}$. With this trick, one finds
\begin{align}
\begin{split}
    \alpha_{\mathrm{tot}}&=\frac{x_{14}a_{23}-x_{13}x_{24}+\pi^{-2}}{2}+\frac{G(x_{14}-x_{23})\pi^{-1}}{2L}\\
    &\quad +\frac{(x_{14}-x_{23})^2+(x_{13}+x_{24})^2}{4L^2}\,,\label{eq:alpha2}
\end{split}
\end{align}
where the variables $x_{ij}$ represent unconstrained entries in the block $\Omega^{(1)}_{AB,A'B'}$. In order to find the asymptotics of EoP, we need to minimize $\alpha_{\mathrm{tot}}$ over these parameters to find the smallest possible EE $S_{AA'}$. Due to the fact that~\eqref{eq:alpha2} is quadratic in $x_{ij}$, we can calculate this valua analytically as
\begin{align}\label{eq:alphatotEoPFerm}
    \alpha_{\mathrm{tot}}&=\frac{1}{8+2\pi^2}\approx 0.03605\,.
\end{align}
Expanding $S_{AA'}\sim \sum_i(\log{2}-\tfrac{\lambda_i}{2})$ through $\lambda_i$ up to second order in $\epsilon_{\sfrac{1}{2}}$ based on~\eqref{eq:symp-eig} allows us to also find the offset analytically, namely we have
\begin{align}
    S_{AA'}=\epsilon_{\sfrac{1}{2}}^2\left(\alpha_{\mathrm{tot}}\log(\epsilon_{\sfrac{1}{2}}^{-2})+\alpha_{\mathrm{tot}} \log{\tfrac{2e}{\alpha_{\mathrm{tot}}}}\right)\,.
\end{align}
Combining this with the result from~\eqref{eq:alphatotEoPFerm} gives
\begin{align}
\label{eq:eopFerm11}
    E_P^\text{fer}&\sim \left(\frac{1}{8+2\pi^2} \log(\epsilon_{\sfrac{1}{2}}^{-2})+ \frac{\log 2e (8+2\pi^2)}{8+2\pi^2} \right)\, \epsilon_{\sfrac{1}{2}}^2 \nonumber\\
    &\sim \left(0.0361\log\left(\tfrac{d}{w}\right)^2 +0.181\right)\, \left(\tfrac{w}{d}\right)^2  \ ,
\end{align}
which agrees with the form~\eqref{eq.SAAp} in the main text. Note that the simplicity of Gaussian states allowed us to even find the analytical form of the constant offset. The accuracy of this analytical prediction was tested numerically, for which we presented the results in Fig.~\ref{fig:MI_EoP} in the main text.

\mysection{EoP for spins} In the limit of an infinite distance between the two single site subsystems, we purify~\eqref{eq:rhoAB} by the state $\ket{\psi^{(0)}}$ with Schmidt decomposition
\begin{align}
\hspace{-7 pt}    \ket{\psi^{(0)}}=\sqrt{D}\ket{\downarrow\downarrow\downarrow\downarrow}+\sqrt{E}(\ket{\uparrow\downarrow\uparrow\downarrow}+\ket{\downarrow\uparrow\downarrow\uparrow})+\sqrt{F}\ket{\uparrow\uparrow\uparrow\uparrow},\label{eq:schmidt}
\end{align}
\normalsize
where the convention for factors ordering in the purification is $ABA'B'$. Note that in this analysis we assume that a minimal purification from two to four spin degrees of freedom suffices and we will subsequently provide supporting numerical evidence and an additional discussion.

Moving on, we \emph{supplement} this purification with finite distance corrections up to second order in $\epsilon_{\sfrac{1}{8}}$ as
\begin{align}
    \ket{\psi}\sim \ket{\psi^{(0)}}+\epsilon_{\sfrac{1}{8}} \ket{\psi^{(1)}}+\tfrac{1}{2}\epsilon_{\sfrac{1}{8}}^2 \ket{\psi^{(2)}}\,.
\end{align}
\normalsize
We will optimize over $\ket{\psi^{(1)}}$ and $\ket{\psi^{(2)}}$ subject to the normalization constraint $\braket{\psi|\psi}=1$ order by order in $\epsilon_{\sfrac{1}{8}}$. We further require $\rho^{(1)}=\ket{\psi^{(0)}}\bra{\psi^{(1)}}+\ket{\psi^{(1)}}\bra{\psi^{(0)}}$ and $\rho^{(2)}=\ket{\psi^{(0)}}\bra{\psi^{(2)}}+\ket{\psi^{(2)}}\bra{\psi^{(0)}}+2\ket{\psi^{(1)}}\bra{\psi^{(1)}}$ to satisfy
\begin{align}
    \rho^{(1)}_{AB}=\begin{pmatrix}
     &&&C\\
     &&C&\\
     &C&&\\
     C&&&
    \end{pmatrix}\quad\text{and}\quad \rho^{(2)}_{AB}=0\,,\label{eq:nG-constaints2}
\end{align}
which follows from\ \eqref{eq:rhoAB}. We expand the first order perturbation as
\begin{align}
    \ket{\psi^{(1)}}=C\sum^{16}_{i=1} z_i\ket{\phi_i}
\end{align}
where $z_i=x_i+\ii y_i$ and $\ket{\phi_i}$ is the basis of $\mathcal{H}_{ABA'B'}$ ordered as $(\ket{\downarrow\downarrow\downarrow\downarrow},\ket{\uparrow\downarrow\downarrow\downarrow},\ket{\downarrow\uparrow\downarrow\downarrow},\ket{\uparrow\uparrow\downarrow\downarrow},\dots,\ket{\uparrow\uparrow\uparrow\uparrow})$. We then need to implement the condition~\eqref{eq:affine-lin} in the main text based on~\eqref{eq:nG-constaints2} together with the normalization constraint~\eqref{eq:constraint1} in the main text. We solve these affine linear constraints by the replacements $x_1=x_6=x_{11}=x_{16}=0$, $z_{5}=-\sqrt{\frac{E}{D}}z_2^*$, $z_{9}=-\sqrt{\frac{E}{D}}z_3^*$, $z_{13}=\frac{1}{\sqrt{D}}-\sqrt{\frac{F}{D}}z_4^*$, $z_{15}=\sqrt{\frac{F}{E}}z_{12}^*$, $z_{10}=\frac{1}{\sqrt{E}}-z_{7}^*$, $z_{14}=-\sqrt{\frac{F}{E}}z_8$.
We can then compute $\alpha_{\mathrm{tot}}$ according to~\eqref{eq:alpha} in the main text as quadratic polynomial in terms of the remaining free variables $z_i$ which leads to the rather involved expression
\begin{widetext}
\begin{align}
\begin{split}
    \hspace{-3mm}\frac{\alpha_{\mathrm{tot}}}{C^2}&
    =\tfrac{(\pi -2)^2 y_1^2}{4 \pi ^2}-\tfrac{2 \sqrt{\pi ^2-4} x_3 x_8}{2+\pi }-\tfrac{2 \sqrt{\pi ^2-4} x_2 x_{12}}{2+\pi }-\tfrac{4 (\pi -2) \pi  x_4}{(2+\pi )^2}-\tfrac{4 \pi  x_7}{\sqrt{\pi ^2-4}}+\tfrac{(\pi -2) \left(-\sqrt{\pi ^2-4} y_6-\sqrt{\pi ^2-4} y_{11}+(2+\pi ) y_{16}\right) y_1}{2 \pi ^2}\\
    &\quad+\tfrac{\left(\pi ^2-4\right) y_6^2}{4 \pi ^2}+\tfrac{\left(\pi ^2-4\right) y_{11}^2}{4 \pi ^2}+\left(\tfrac{1}{4}+\tfrac{1}{\pi ^2}+\tfrac{1}{\pi }\right) y_{16}^2+\tfrac{(2+\pi ) y_6 \left((\pi -2) y_{11}-\sqrt{\pi ^2-4} y_{16}\right)}{2 \pi ^2}-\tfrac{(2+\pi ) \sqrt{\pi ^2-4} y_{11} y_{16}}{2 \pi ^2}\\
    &\quad-\tfrac{2 \sqrt{\pi ^2-4} y_3 y_8}{2+\pi }-\tfrac{2 \sqrt{\pi ^2-4} y_2 y_{12}}{2+\pi }+\tfrac{(\pi -2) |z_2|^2}{2+\pi }+\tfrac{(\pi -2) |z_3|^2}{2+\pi }+\tfrac{2 \left(4+\pi ^2\right) |z_4|^2}{(2+\pi )^2}+2 |z_7|^2+|z_8|^2+|z_{12}|^2+\tfrac{8 \pi ^3}{(\pi -2) (2+\pi )^2}\,.
\end{split}
\end{align}
\end{widetext}
In order to find the EoP, we need to minimize over the $z_i$ to find the smallest possible value $\alpha_{\mathrm{tot}}$, which can be done \emph{analytically} and leads to
\begin{align}
\label{eq.alphatot}
    \alpha_{\mathrm{tot}}=\frac{4\pi^4C^2}{\pi^4-16}\approx 0.12445\,.
\end{align}
The non-vanishing $\alpha_{\mathrm{tot}}$ shows that the resulting EoP obtained from~\eqref{eq.SAAp} again has the form 
\begin{align}
\label{EQ_EOP_SPIN_W1}
    E_P^\text{spin} &\sim \left(  \frac{4\pi^4C^2}{\pi^4-16} \log(\epsilon_{\sfrac{1}{8}}^{-2}) + \mathrm{const} \right) \epsilon_{\sfrac{1}{8}}^{2} \nonumber\\
    &\sim \left( 0.124 \log\sqrt{\frac{d}{w}} + 0.440 \right) \sqrt{\frac{w}{d}} \ ,
\end{align}
which, as in the fermion case, exhibits a leading-order long-distance behavior enhanced with respect to that of MI~\eqref{eq.MIanalytic} by a logarithm of the distance.

When it comes to the \emph{subleading} long-distance behavior encapsulated by~$\Big( \sum_{j>0}\alpha_j(1-\log{\alpha_j}) \Big)$, we would need to extract the individual $\alpha_j$ and optimize over the remaining parameters. While it is plausible this can be also done analytically, we determined the value quoted above numerically, as discussed in the main text. Note that for the free fermion case with $w=\delta$ considered above, we determined this term analytically in terms of~$\alpha_{\mathrm{tot}}$.

\mysection{RE for fermions} In the Gaussian case of free fermions, our starting point is the following perturbative expansion of the reduced density matrix $\rho_{AB}$ of a system of $1+1$ fermions in the large $d$ separation, $\rho_{AB}\sim\rho^{(0)}_{A}\otimes\rho^{(0)}_{B}+\epsilon_{1/2}\rho^{(1)}_{AB}$ given by~\eqref{eq:rho-fermions}. We similarly construct the canonical purification of~\eqref{eq:rho-fermions} via $\ket{\sqrt{\rho_{AB}}}=\sum_{i}\sqrt{e_{i}}\ket{e_{i}}\otimes\ket{e_{i}}=\ket{\psi^{(0)}}+\epsilon_{\sfrac{1}{2}}\, \ket{\psi^{(1)}}$ where $\rho_{AB}\ket{e_{i}}=e_{i}\ket{e_{i}}$. Note that in contrast with fermionic MI and EoP, we do not need to phrase our computation of RE in terms of the covariance matrix formalism since we can construct the canonical purification $\ket{\sqrt{\rho_{AB}}}$exactly for the given form of the initial reduced density matrix~$\rho_{AB}$. 

In this case, the first-order perturbation $\ket{\psi^{(1)}}$ is simply given by
\begin{align}\label{EQ_PSI1_FERMION}
    \ket{\psi^{(1)}}=\tfrac{1}{2\pi}(\ket{\phi_{4}}+\ket{\phi_{13}}) \, ,
\end{align}
with the same ordering of the basis $\ket{\phi_{i}}$ as in the previous case. From the canonical purification's density matrix $\rho:=\ket{\sqrt{\rho_{AB}}}\bra{\sqrt{\rho_{AB}}}$ we consider a restriction to subsystems $AA'$ given by the reduced density matrix $\rho_{AA'}=\textrm{tr}_{BB'}(\rho)$ which has the perturbative expansion $\rho_{AA'}=\rho^{(0)}_{A}\otimes\rho^{(0)}_{A'}+\epsilon^{2}_{\sfrac{1}{2}}\rho^{(2)}_{AA'}$ explicitly given by
\begin{align}
\label{EQ_RHO_AA_FERMION}
   \rho_{AA'} \sim \rho^{(0)}_{AA'}+\tfrac{1}{2}\epsilon^{2}_{\sfrac{1}{2}}\rho^{(2)}_{AA'}=
   \begin{pmatrix}
  \tilde{G}_{1} &  &  & \tilde{H} \\
  & \tilde{J} &  & \\
  &  &   \tilde{J}&  \\
 \tilde{H} &  &  & \tilde{G}_{2}
    \end{pmatrix} ,
\end{align}
where $\tilde{G}_{1}=\frac{\pi+2}{2\pi}-\frac{\epsilon^{2}_{\sfrac{1}{2}}}{4\pi^{2}}$, $\tilde{G}_{2}=\tilde{G}_{1}-\frac{2}{\pi}$, $\tilde{H}=\frac{\sqrt{\pi^{2}-4}}{2\pi}-\frac{\sqrt{\pi^{2}-4}\epsilon^{2}_{\sfrac{1}{2}}}{4\pi(\pi^{2}-4)}$, and $\tilde{J}=\frac{\epsilon^{2}_{\sfrac{1}{2}}}{4\pi^{2}}$. We once again compute the trace of the square of~\eqref{EQ_RHO_AA_FERMION} according to~\eqref{eq:alpha} in the main text from which we obtain
\begin{align}\label{EQ_APLHA_TOT_RE_FERMION}
    \alpha_{\mathrm{tot}}=\frac{1}{2\pi^{2}}\approx 0.051 \, .
\end{align}
which also shows that the reflected entropy $S_{R}(\rho_{AB})=S_{AA'}(\rho)$ of the fermionic subsystem also exhibits a logarithmic enhancement of the power law decay for $w=\delta$ given by
\begin{align}\label{EQ_RE_FERMION_W1}
    S_{R}^\text{fer}(\rho_{AB}) &\sim 
    \left( \frac{1}{2\pi^{2}} \log\epsilon_{\sfrac{1}{2}}^{-2} + \frac{1 + \log(4\pi^2)}{2\pi^2} \right) \epsilon_{\sfrac{1}{2}}^2 \nonumber  \\
    & \sim\left(0.051\log\left(\frac{d}{w}\right)^2+0.237\right) \left(\frac{w}{d}\right)^{2}\, ,
\end{align}
where we also computed the constant term in~\eqref{EQ_RE_FERMION_W1} from the eigenvalues of~\eqref{EQ_RHO_AA_FERMION} according to~\eqref{eq:alpha} in the main text.

\mysection{RE for spins} For the Ising spin case, we now describe the detailed computation of the reflected entropy RE for $w=\delta$ in the large $d$ limit just as for fermions. The reduced density matrix for a spin system of $1+1$ sites in the large $d$ limit can again be computed according to~\eqref{eq.rhogeneral}, \ie $\rho_{AB} \sim \rho^{(0)}_{A} \otimes \rho^{(0)}_{B}+\epsilon_{\sfrac{1}{8}}\, \rho^{(1)}_{AB}+\ldots$, yielding~\eqref{eq:rhoAB}.

We now construct the canonical purification of~\eqref{eq:rhoAB} via $\ket{\sqrt{\rho_{AB}}}=\sum_{i}\sqrt{e_{i}}\ket{e_{i}}\otimes\ket{e_{i}}=\ket{\psi^{(0)}}+\epsilon_{\sfrac{1}{8}}\ket{\psi^{(1)}}$ for $\rho_{AB}\ket{e_{i}}=e_{i}\ket{e_{i}}$ and where the eigenvalues $e_{i}$ are defined in~\eqref{eq:EigenSpinRhoAB}. In this case, the first order perturbation $\ket{\psi^{(1)}}$ is given by
\begin{align} \label{EQ_PSI1_ISING}
\begin{split}
    \ket{\psi^{(1)}}&= \tfrac{\pi}{\sqrt{\pi^{2}-4}}(\ket{\phi_{7}}+\ket{\phi_{10}})+\ket{\phi_{4}}+\ket{\phi_{13}}\, ,
    \end{split}
\end{align}
where the states $\ket{\phi_{i}}=\ket{\phi_{i}}_{ABA'B'}$ form an orthonormal basis for the purified Hilbert space $\mathcal{H}_{ABA'B'}$ ordered as $(\ket{\downarrow\downarrow\downarrow\downarrow},\ket{\uparrow\downarrow\downarrow\downarrow},\ket{\downarrow\uparrow\downarrow\downarrow},\ket{\uparrow\uparrow\downarrow\downarrow},\dots,\ket{\uparrow\uparrow\uparrow\uparrow})$. From the canonical purification's density matrix $\rho:=\ket{\sqrt{\rho_{AB}}}\bra{\sqrt{\rho_{AB}}}$ we consider a restriction to subsystems $AA'$ given by the reduced density matrix $\rho_{AA'}=\textrm{tr}_{BB'}(\rho)$ which has the perturbative expansion $\rho_{AA'}=\mathrm{tr}_{BB'}(\ket{\psi^{(0)}}\bra{\psi^{(0)}})+\epsilon^{2}_{\sfrac{1}{8}}(2\mathrm{tr}_{BB'}(\ket{\psi^{(1)}}\bra{\psi^{(1)}}))/2$ explicitly given by
\begin{align}
\label{EQ_RHO_AA_SPIN}
   \rho_{AA'} \sim\rho^{(0)}_{AA'}+\tfrac{1}{2}\epsilon^{2}_{\sfrac{1}{8}}\rho^{(2)}_{AA'}=
   \begin{pmatrix}
  \tilde{A}_{1} &  &  & \tilde{F} \\
  & \tilde{B} & \tilde{E} & \\
  & \tilde{E} & \tilde{B} &  \\
 \tilde{F} &  &  & \tilde{A}_{2}
    \end{pmatrix} \, ,
\end{align}
where $\tilde{A}_{1}=\frac{\pi+2}{2\pi}-\frac{2(\pi^{2}-2)C^{2}\epsilon^{2}_{\sfrac{1}{8}}}{\pi^{2}-4},\tilde{A}_{2}=\tilde{A}_{1}-\frac{2}{\pi},\tilde{B}=\frac{2(\pi^{2}-2)C^{2}\epsilon^{2}_{\sfrac{1}{8}}}{(\pi^{2}-4)},\tilde{E}=\frac{2\pi C^{2}\epsilon^{2}_{\sfrac{1}{8}}}{\sqrt{\pi^{2}-4}},\tilde{F}=\frac{\sqrt{\pi^{2}-4}}{2\pi}-\frac{2\pi(\pi^{2}-2)C^{2}\epsilon^{2}_{\sfrac{1}{8}}}{(\pi^{2}-4)^{3/2}}$, where the coefficient $C$ is defined as in~\eqref{eq:Cdefinition}. From here we follow the strategy of the main text and compute the trace of the the square of~\eqref{EQ_RHO_AA_SPIN} according to~\eqref{eq:alpha}.
In this case, we find a value of $\alpha_{\mathrm{tot}}$ computed via~\eqref{eq:alpha} to be
\begin{align}
  \alpha_{\mathrm{tot}}=\frac{4C^{2}(\pi^{2}-2)}{\pi^{2}-4}\approx0.139 \,.\label{eq:alpha-RE-ISING}
\end{align}
As a consequence, the large $d$ leading behaviour of the reflected entropy $S_{R}(\rho_{AB}):=S_{AA'}(\rho)$ exhibits a non trivial logarithmic enhancement of the power law decay according to~\eqref{eq.SAAp} and where the constant contribution can be computed from the eigenvalues of~\eqref{EQ_RHO_AA_SPIN} leading to a reflected entropy $S_{R}$ of the Ising subsystem for $w=\delta$ of
\begin{align}\label{EQ_RE_SPIN_W1}
  S_{R}^\text{spin}(\rho_{AB}) &\sim 
  \left( \frac{4C^2\pi^4}{\pi^4 - 16} \log\epsilon_{\sfrac{1}{8}}^{-2} + \mathrm{const} \right)\epsilon_{\sfrac{1}{8}}^2 \nonumber\\
  &\sim \left(0.139\log\sqrt{\frac{d}{w}} + 0.425 \right) \sqrt{\frac{w}{d}}\, .
\end{align}
The constant term is again determined numerically in the main text.

\mysection{Numerical approach and asymmetric purifications}
Our numerical methods are based on
~[29,~70,~71], which outline the construction of an efficient algorithm for local optimization over Gaussian states, based on a gradient descent approach exploiting the natural Lie group parametrization of the state manifolds. Our numerical results are obtained using an adaptation of this algorithm to the non-Gaussian case of interest.

To compute the EoP as given in~\eqref{eq.EoPdef}, we minimise EE $S$ over the manifold $\mathcal{M}$ of purified state density matrices. We first purify our initial mixed density matrix to a $2^N$-dimensional pure $\rho_\id$ via the Schmidt decomposition. Here, $N=\sum_X N_X$ with $N_X$ denoting the physical degrees of freedom in subsystem $X$. We parametrize elements $\rho_U\in\mathcal{M}$ by transformations $U=\id\otimes\widetilde{U}$ with $\widetilde{U}\in\mathrm{U}(2^{N_{A'}+N_{B'}})$, so that $\rho_U=U\rho_\id U^{-1}$. The tensor product signifies that $U$ only acts non-trivially on degrees of freedom in $A'$ and $B'$. We then optimize by performing iterative steps along directions in $\mathcal{M}$ which locally minimize $S_{AA'}$~[29,70],
\begin{align}
    \label{eq.single_step}
    U_{n+1}=U_n\mathrm{e}^{t K_n}\,.
\end{align}
Here, $K_n=\sum_\mu\mathcal{F}^\mu(U_n)\Xi_\mu/||\mathcal{F}||^2$ and $\mathcal{F}^\mu:~\mathcal{M}\to\mathbb{R}$ is the gradient descent vector field
\begin{align}
    \mathcal{F}^\mu(U)=-\frac{\partial}{\partial s}S(U\mathrm{e}^{s\Xi_\mu}\rho_\id\mathrm{e}^{-s\Xi_\mu}U^{-1})|_{s=0}
\end{align}
with $\{\Xi_\mu\}$ as basis of $\mathfrak{u}(2^{N_{A'}+N_{B'}})$. We choose $U_0=\id$ and we pick $0<t<1$ in such a way that the value of $S_{AA'}$ decreases with successive steps.

\begin{table}[t]
    \centering
    \renewcommand{\arraystretch}{1.25}
\begin{tabular}{@{} @{\hskip 7pt}c @{\hskip 7pt}c @{\hskip 7pt}| @{\hskip 7pt} c @{\hskip 7pt}| c @{\hskip 15pt} c c @{\hskip 15pt} c c c @{}
}
\toprule
&& & \multicolumn{6}{c}{$N_{A'}+N_{B'}$} \\
&& & $1+1$ & $1+2$ & $2+1$ & $1+3$ & $2+2$ & $3+1$
\\

\colrule
\parbox[t]{2mm}{\multirow{8}{*}{\rotatebox[origin=c]{90}{$N_A+N_B$}}} & \parbox[t]{2mm}{\multirow{4}{*}{\rotatebox[origin=c]{90}{$1+1$}}} & $d=\delta\;$ & \cellcolor{yellow!35}{$0.382$} & \cellcolor{yellow!10}{$0.382$} & \cellcolor{yellow!10}{$0.382$} & \cellcolor{yellow!10}{$0.382$} & \cellcolor{yellow!10}{$0.382$} & \cellcolor{yellow!10}{$0.382$} 
\\
&& $d=2\delta$ & \cellcolor{yellow!35}{$0.333$} & \cellcolor{yellow!10}{$0.333$} & \cellcolor{yellow!10}{$0.333$} & \cellcolor{yellow!10}{$0.333$} & \cellcolor{yellow!10}{$0.333$} & \cellcolor{yellow!10}{$0.333$} 
\\
&& $d=3\delta$ & \cellcolor{yellow!35}{$0.306$} & \cellcolor{yellow!10}{$0.306$} & \cellcolor{yellow!10}{$0.306$} & \cellcolor{yellow!10}{$0.306$} & \cellcolor{yellow!10}{$0.306$} & \cellcolor{yellow!10}{$0.306$} 
\\
&& $d=2\delta$ & \cellcolor{yellow!35}{$0.292$} & \cellcolor{yellow!10}{$0.292$} & \cellcolor{yellow!10}{$0.292$} & \cellcolor{yellow!10}{$0.292$} & \cellcolor{yellow!10}{$0.292$} & \cellcolor{yellow!10}{$0.292$} 
\\\hhline{*{1}{~}*{8}{-}}

& \parbox[t]{2mm}{\multirow{4}{*}{\rotatebox[origin=c]{90}{$1+2$}}} & $d=\delta\;$ & \multirow{4}{*}{n.a.} & \cellcolor{yellow!35}{$0.412$} & $0.438$ & \cellcolor{yellow!10}{$0.412$} & \cellcolor{yellow!10}{$0.412$} & $0.440$ 
\\
&& $d=2\delta$ && \cellcolor{yellow!35}{$0.368$} & $0.412$ & \cellcolor{yellow!10}{$0.368$} & \cellcolor{yellow!10}{$0.368$} & $0.415$ 
\\
&& $d=3\delta$ && \cellcolor{yellow!35}{$0.345$} & $0.394$ & \cellcolor{yellow!10}{$0.345$} & \cellcolor{yellow!10}{$0.345$} & $0.398$
\\
&& $d=4\delta$ && \cellcolor{yellow!35}{$0.335$} & $0.385$ & \cellcolor{yellow!10}{$0.335$} & \cellcolor{yellow!10}{$0.335$} & $0.389$
\\\botrule

\end{tabular}
    \ccaption{Numerical evidence for optimality of certain minimal purifications}{The table shows the values of the optimization for different choices of the system dimensions and of $d$. The true EoP values (the minimum optimization values) are highlighted in yellow, with the darker shade indicating the lowest-dimensional purification for which the EoP is obtained.}
\label{tab:non-minimal purifications}
\end{table}

The $\{\Xi_\mu\}$ span the tangent space at $U=\id$ and, due to the left-invariance of the Riemannian metric on $\mathcal{M}$, form  orthonormal bases for the tangent spaces at all other points in $\mathcal{M}$, too, where $\Xi_\mu$ is identified with the tangent vector to the curve $\gamma(s)=U\mathrm{e}^{s\Xi_\mu}$ at $\gamma(0)$~[70].
This saves us having to re-evaluate the matrix representation of the metric at each step, as we would have to if we had chosen a coordinate parametrisation of $\mathcal{M}$. While this makes our algorithm more efficient than a naive gradient descent, the numerically accessible range is still highly limited: since $N_{A'}+N_{B'}\geq N_A+N_B$, the dimension of $\mathcal{M}$ is at least $\mathrm{dim}\mathfrak{u}(2^{N_{A'}+N_{B'}})=2^{2N_{A'}+2N_{B'}}-1$ and~\eqref{eq.single_step} requires exponentiation of at least $2^{(N_A+N_B)}\times2^{(N_A+N_B)}$ matrices, with a typical step count of several hundred. This becomes extremely slow on a powerful desktop computer for $N_{A'}+N_{B'}\geq 5$. For the symmetric purifications in the main text this corresponds with $w>2\,\delta$, which explains the regime we were able to explore.

Given this limitation on our numerical capabilities, it is instructive to ask whether an optimization over \textsl{minimal purifications} corresponding with $N_{A'}+N_{B'}=N_A+N_B$ yields the true minimum of EE~\textendash~not least because for large systems this becomes the only numerically viable choice. A natural follow-up question is whether among the choices of minimal purifications, the intuitive choice of $N_{A'}=N_A$ and $N_{B'}=N_B$ suffices to reach the true minimum defined as EoP. More pertinently, we might ask whether it is even possible to reach the true minimum with a minimal purification for which $N_{A'}\neq N_A$ and $N_{B'}\neq N_B$. In~[70], 
a combination of numerical and analytical evidence was provided to show that the answer to this question is in affirmative for Gaussian states. While limited by the greater numerical challenge in the non-Gaussian case, we present similar numerical evidence in Table~\ref{tab:non-minimal purifications} to show that the same may be said for our model: the true minimum can only be reached if $N_{A'}\geq N_{A}$ and $N_{B'}\geq N_{B}$, which indicates that the lowest-dimensional purification for which the EoP can be obtained is the minimal purification with $N_{A'}=N_A$ and $N_{B'}=N_B$.

\bibliography{references} 
\end{document}